\documentclass[12pt]{article}
\usepackage{graphics,psfrag}
\usepackage{euscript,array,cite}

\newcommand{\ba}{\begin{eqnarray}}
\newcommand{\ea}{\end{eqnarray}}
\newcommand{\nn}{\nonumber}
\newcommand{\cZ}{{\mathcal{Z}}}
\newcommand{\cA}{{\mathcal{A}}}

\newcounter{multieqs}

\newcommand{\be}{\begin{equation}}
\newcommand{\ee}{\end{equation}}
\newcommand{\eq}[1]{(\ref{#1})}

\newcommand{\bm}[1]{\mbox{\boldmath $#1$}}
\newcommand{\rf}[1]{(\ref{#1})}

\def\bd{\begin{document}}
\def\ed{\end{document}}
\def\nn{\nonumber}
\def\bea{\begin{eqnarray}}
\def\eea{\end{eqnarray}}
\let\bm=\bibitem
\let\la=\label

\def\npb#1#2#3{Nucl. Phys. {\bf{B#1}} #3 (#2)}
\def\plb#1#2#3{Phys. Lett. {\bf{#1B}} #3 (#2)}
\def\prl#1#2#3{Phys. Rev. Lett. {\bf{#1}} #3 (#2)}
\def\prd#1#2#3{Phys. Rev. {D \bf{#1}} #3 (#2)}
\def\cmp#1#2#3{Comm. Math. Phys. {\bf{#1}} #3 (#2)}
\def\cqg#1#2#3{Class. Quantum Grav. {\bf{#1}} #3 (#2)}
\def\nppsa#1#2#3{Nucl. Phys. B (Proc. Suppl.) {\bf{#1A}}#3 (#2)}
\def\ap#1#2#3{Ann. of Phys. {\bf{#1}} #3 (#2)}
\def\ijmp#1#2#3{Int. J. Mod. Phys. {\bf{A#1}} #3 (#2)}
\def\rmp#1#2#3{Rev. Mod. Phys. {\bf{#1}} #3 (#2)}
\def\mpla#1#2#3{Mod. Phys. Lett. {\bf A#1} #3 (#2)}
\def\jhep#1#2#3{J. High Energy Phys. {\bf #1} #3 (#2)}
\def\atmp#1#2#3{Adv. Theor. Math. Phys. {\bf #1} #3 (#2)}

\def\N{{\cal N}}
\def\sst{\scriptscriptstyle}
\def\thetabar{\bar\theta}
\def\Tr{{\rm Tr}}
\def\one{\mbox{1 \kern-.59em {\rm l}}}

\def\a{\alpha}      \def\da{{\dot\alpha}}  \def\dA{{\dot A}}
\def\b{\beta}       \def\db{{\dot\beta}}  
\def\g{\gamma}  \def\G{\Gamma}  \def\dc{{\dot\gamma}}  
\def\d{\delta}  \def\D{\Delta}  \def\ddt{\dot\delta}  
\def\e{\epsilon}        \def\ve{\varepsilon}  
\def\f{\phi}    \def\F{\Phi}    \def\vvf{\f}  
\def\h{\eta}  
\def\k{\kappa}  
\def\l{\lambda} \def\L{\Lambda}  
\def\m{\mu} \def\n{\nu}  
\def\o{\omega}  
\def\p{\pi} \def\P{\Pi}  
\def\r{\rho}  
\def\s{\sigma}  \def\S{\Sigma}  
\def\t{\tau}  
\def\th{\theta} \def\Th{\Theta} \def\vth{\vartheta}  
\def\X{\Xeta}  
\def\z{\zeta}  

\def\na{\nabla}  

\def\cA{{\cal A}} \def\cB{{\cal B}} \def\cC{{\cal C}}  
\def\cD{{\cal D}} \def\cE{{\cal E}} \def\cF{{\cal F}}  
\def\cG{{\cal G}} \def\cH{{\cal H}} \def\cI{{\cal I}}  
\def\cJ{{\cal J}} \def\cK{{\cal K}} \def\cL{{\cal L}}  
\def\cM{{\cal M}} \def\cN{{\cal N}} \def\cO{{\cal O}}  
\def\cP{{\cal P}} \def\cQ{{\cal Q}} \def\cR{{\cal R}}  
\def\cS{{\cal S}} \def\cT{{\cal T}} \def\cU{{\cal U}}  
\def\cV{{\cal V}} \def\cW{{\cal W}} \def\cX{{\cal X}}  
\def\cY{{\cal Y}} \def\cZ{{\cal Z}}

\def\ua{\underline{\alpha}}  
\def\uc{\underline{\phantom{\alpha}}\!\!\!\gamma}  
\def\um{\underline{\mu}}  
\def\ud{\underline\delta}  
\def\ue{\underline\epsilon}  
\def\una{\underline a}\def\unA{\underline A}  
\def\unb{\underline b}\def\unB{\underline B}  
\def\unc{\underline c}\def\unC{\underline C}  
\def\und{\underline d}\def\unD{\underline D}  
\def\une{\underline e}\def\unE{\underline E}  
\def\unf{\underline{\phantom{e}}\!\!\!\! f}\def\unF{\underline F}  
\def\unm{\underline m}\def\unM{\underline M}  
\def\unn{\underline n}\def\unN{\underline N}  
\def\unp{\underline{\phantom{a}}\!\!\! p}\def\unP{\underline P}  
\def\unq{\underline{\phantom{a}}\!\!\! q}  
\def\unQ{\underline{\phantom{A}}\!\!\!\! Q}  
\def\unH{\underline{H}}  
  
\def\As {{A \hspace{-6.4pt} \slash}\;}  
\def\bs {{b \hspace{-6.4pt} \slash}\;}  
\def\Ds {{D \hspace{-6.4pt} \slash}\;}  
\def\ds {{\del \hspace{-6.4pt} \slash}\;}  
\def\ss {{\s \hspace{-6.4pt} \slash}\;}  
\def\ks {{ k \hspace{-6.4pt} \slash}\;}  
\def\ps {{p \hspace{-6.4pt} \slash}\;}   
\def\xs {{x \hspace{-6.4pt} \slash}\;}  
\def\pas {{{p_1} \hspace{-6.4pt} \slash}\;}  
\def\pbs {{{p_2} \hspace{-6.4pt} \slash}\;}   
\def\cFs {{{\cal F} \hspace{-6.4pt} \slash}\;}

\def\Dh{\hat{D}}
\def\Gh{\hat{G}}
\def\Fh{\hat{F}}
\def\Ph{\hat{P}}
\def\Rh{\hat{R}}
\def\Vh{\hat{V}}  
\def\Xh{\hat{X}}  
\def\Yh{\hat{Y}} 
\def\Ah{\hat{A}}
\def\Bh{\hat{B}}
\def\Rh{\hat{R}}
\def\Ch{\hat{C}}
\def\Psih{\hat{\Psi}}

\def\ah{\hat{a}}
\def\gh{\hat{g}} 
\def\hh{\hat{h}}
\def\uh{\hat{u}}  
\def\xh{\hat{x}}  
\def\yh{\hat{y}}  
\def\ph{\hat{p}}  
\def\xih{\hat{\xi}}  
\def\chih{\hat{\chi}}  
\def\ih{\hat{i}} 
\def\jh{\hat{j}} 
\def\kh{\hat{k}}

\def\psit{\tilde{\psi}}  
\def\Psit{\tilde{\Psi}}   
\def\Psibt{\tilde{\bar{Psi}}}  

\def\st{\tilde{\sigma}}  

\def\Phit{\tilde{\Phi}}   
\def\Phitb{\overline{\tilde{Phi}}}  
\def\tht{\tilde{\th}}  
\def\lt{\tilde{\l}}
\def\chit{\tilde{\chi}}   
\def\phit{\tilde{\phi}} 

\def\At{\tilde{A}}
\def\Bt{\tilde{B}}
\def\Ct{\tilde{C}}
\def\Dt{\tilde{D}}
\def\Ft{\tilde{F}}
\def\Qt{\tilde{Q}}  
\def\Rt{\tilde{R}}  
\def\Mt{\tilde{M }}  
\def\Nt{\tilde{N}}   
\def\St{\tilde{S}}
\def\Vt{\tilde{V}}
\def\Xt{\tilde{X}} 
\def\at{\tilde{a}}
\def\ct{\tilde{c}}   
\def\htt{\tilde{h}} 
\def\ft{\tilde{f}}
\def\gt{\tilde{g}}
\def\pt{\tilde{p}}  
\def\qt{\tilde{q}}  
\def\vt{\tilde{v}}  
\def\nt{\tilde{n}}  
\def\ut{\tilde{u}}  
\def\wt{\tilde{w}}  
\def\zt{\tilde{z}} 
\def\xt{\tilde{x}} 
\def\yt{\tilde{y}} 
\def\Psit{\tilde{\Psi}}
\def\vphit{\tilde{\varphi}} 
\def\Lt{\tilde{\L}}   

\def\delb{\bar{\partial}}  
\def\thb{\bar{\theta}}
\def\mub{\bar{\mu}}
\def\lamb{\bar{\l}}
\def\psib{\bar{\psi}}
\def\sb{\bar{\sigma}}
\def\xib{\bar{\xi}}
\def\chib{\bar{\chi}}

\def\Phib{\bar{\Phi}}
\def\Lamb{\bar{\Lambda}}
\def\Sb{{\overline \Sigma}}
\def\cb{\bar{c}}
\def\qb{\bar{q}}
\def\wb{\bar{w}}
\def\ub{\bar{u}}
\def\zb{{\bar{z}}}
\def\Qb{{\bar Q}}
\def\Psib{\bar{\Psi}}

\def\Ab{{\overline A}} \def\Bb{{\overline B}} \def\Cb{{\overline C}}  
\def\Db{{\overline D}} \def\Eb{{\overline E}} \def\Fb{{\overline F}}  
\def\Gb{{\overline G}} \def\Hb{{\overline H}} \def\Ib{{\overline I}}  
\def\Jb{{\overline J}} \def\Kb{{\overline K}} \def\Lb{{\overline L}}  
\def\Mb{{\overline M}} \def\Nb{{\overline N}} \def\Ob{{\overline O}}  
\def\Pb{{\overline P}}  \def\Rb{{\overline R}}  
 \def\Tb{{\overline T}} \def\Ub{{\overline U}}  
\def\Vb{{\overline V}} \def\Wb{{\overline W}} \def\Xb{{\overline X}}  
\def\Yb{{\overline Y}} \def\Zb{{\overline Z}}  

\def\fb{{\overline f}}
\def\gb{{\overline g}}
\def\mb{{\overline m}}
\def\lb{{\overline l}}
\def\yb{{\overline y}}
\def\eb{\bar{\e}} 

\def\bk{{\bf k}}  
\def\bl{{\bf l}}  
\def\bp{{\bf p}}  
\def\bq{{\bf q}}  
\def\br{{\bf r}}
\def\bt{{\bf t}}
\def\bu{{\bf u}}
\def\bv{{\bf v}}
\def\bx{{\bf x}}  
\def\by{{\bf y}}  
\def\bR{{\bf R}}  
\def\bV{{\bf V}}

\def\bone{{\bf 1}}  

\def\va{{\vec a}}
\def\vk{{\vec k}}
\def\vp{{\vec p}}
\def\vq{{\vec q}}
\def\vx{{\vec x}}
\def\vy{{\vec y}}
\def\vu{{\vec u}}
\def\vv{{\vec v}}

\def\vs{{\vec \sigma}}
\def\vtau{{\vec \tau}}

\newcommand{\ov}[1]{\overrightarrow{#1}}
  
\def\d{\delta}\def\D{\Delta}\def\ddt{\dot\delta}  
  
\def\pa{\partial} \def\del{\partial}  
\def\xx{\times}  
\def\uno{\mbox{1 \kern-.59em {\rm l}}}    
  
\def\trp{^{\top}}  
\def\inv{^{-1}}  
\def\dag{{^{\dagger}}}  
\def\pr{^{\prime}}  
  
\def\rar{\rightarrow}  
\def\lar{\leftarrow}  
\def\lrar{\leftrightarrow}  
  
\newcommand{\0}{\,\!}      
\def\one{1\!\!1\,\,}  
\def\im{\imath}  
\def\jm{\jmath}  
  
\newcommand{\tr}{\mbox{tr}}  
\newcommand{\slsh}[1]{/ \!\!\!\! #1}  
  
\def\vac{|0\rangle}  
\def\lvac{\langle 0|}  
  
\def\hlf{\frac{1}{2}}  
\def\ove#1{\frac{1}{#1}}  

\def\Box{\square}  
\def\CC {\mathbb{C}}
\def\RR{\mathbb{R}}
\def\ZZ{\mathbb{Z}}  
\def\bb#1{{\bf #1}}  
\def\bcomment#1{}  
\def\bfhat#1{{\bf \hat{#1}}}  
\def\VEV#1{\left\langle #1\right\rangle}  

\newcommand{\ex}[1]{{\rm e}^{#1}} \def\ii{{\rm i}}  

\newcommand{\lrbrk}[1]{\left(#1\right)}
\newcommand{\sfrac}[2]{{\textstyle\frac{#1}{#2}}}
 
\def\stw{{\sqrt{2}}}

\def\rf {{\rm f}}
\def\ri {{\rm i}}
\def\rs {{\scriptscriptstyle \rm S}}
\def\rt {{\scriptscriptstyle \rm T}}

\def\rQ {{\scriptscriptstyle \rm \cQ}}
\def\rR {{\scriptscriptstyle \rm \cR}}

\def\cQb{{\cal \Qb}}
\def\cRb{{\cal \Rb}}
\def\cWb{{\cal \Wb}}

\def\fd {{\rm N}}
\def\afd {{\overline{\rm N}}}

\def \II {I\hspace{-.1em}I\hspace{.1em}}
\def \IIA {\mbox{\II A\hspace{.2em}}}
\def \IIB {\mbox{\II B\hspace{.2em}}}
\def \gs {g^s}
\def \ls {\lambda^s}

\def \I {{\cal I}}
\def \qs {q\hspace{-.53em}/\hspace{.15em}}
\def \ks {k\hspace{-.53em}/\hspace{.15em}}
\def \YM {{\mbox{\tiny YM}}}
\def \gym {g_{\YM}}

\def \Lc {\L_c}

\begin{document}

\begin{titlepage}

\begin{center}

\hfill UT-08-19
\vskip .5in

\textbf{\Large Truncated Nambu-Poisson Bracket and
Entropy Formula for Multiple Membranes}

\vskip 15 mm

{\large
Chong-Sun Chu$^\natural$\footnote{
email address: chong-sun.chu@durham.ac.uk},
Pei-Ming Ho$^\dagger$\footnote{
e-mail address: pmho@phys.ntu.edu.tw}, 
Yutaka Matsuo$^\ddagger$\footnote{
e-mail address: matsuo@phys.s.u-tokyo.ac.jp} 
and Shotaro Shiba$^\ddagger$\footnote{
e-mail address: shiba@hep-th.phys.s.u-tokyo.ac.jp}
 }\\
\vskip 10mm
{\it
$^\natural$
Centre for Particle Theory and Department of Mathematics, Durham\
University of Durham, DH1 3LE, UK
}\\
\vskip 3mm
{\it
$^\dagger$
Department of Physics and Center for Theoretical Sciences, \\
National Taiwan University, Taipei 10617, Taiwan,
R.O.C.}\\
\vskip 3mm
{\it
$^\ddagger$
Department of Physics, Faculty of Science, University of Tokyo,\\
Hongo 7-3-1, Bunkyo-ku, Tokyo 113-0033, Japan\\
\noindent{ \smallskip }\\
}
\vspace{60pt}
\end{center}
\begin{abstract}

We show that there exists a cut-off version of
Nambu-Poisson bracket which defines a finite dimensional 
Lie 3-algebra.
The algebra still satisfies the fundamental identity
and thus produces $\mathcal{N}=8$ supersymmetric
BLG type equation of motion for multiple M2 branes.
By counting the number of the moduli and the degree of freedom,
we derive an entropy formula which scales as $N^{3/2}$ as expected
for the multiple M2 branes.
\end{abstract}

\end{titlepage}
\setcounter{footnote}{0}

\section{Introduction}

Bagger-Lambert-Gustavsson model \cite{Bagger:2006sk,Bagger:2007jr,
Bagger:2007vi,Gustavsson:2007vu}
gives a prototype construction of an $\cN=8$ superconformal
field theory in 3 dimensions. The construction relies on two
structures of a Lie 3-bracket:
the fundamental identity which is essential to the closure of the
supersymmetry transformations; and an invariant metric which allows one
to convert the equation of motion to an off-shell action. The original
example of BL was based on a  Lie 3-algebra 
$A_4$ with a positive definite metric.  In this case, one may alternatively
define the theory by a Lie algebra $SU(2) \times
SU(2)$ \cite{VanRaamsdonk:2008ft}.  It was then proved 
that this is essentially the only  possible
3-algebra which satisfies all the requirements
\cite{Nagy, Papadopoulos:2008sk,Gauntlett:2008uf} (see also \cite{Ho:2008bn}).

It turned out that one may replace the conditions for Lie 3-algebra
by milder ones.  One possibility is to permit to include a
negative norm generator \cite{Gomis:2008uv,Benvenuti:2008bt,%
Ho:2008ei} (see also \cite{FigueroaO'Farrill:2008zm, 
Bandres:2008kj,Gomis:2008be,Ezhuthachan:2008ch,
Cecotti:2008qs}).
In this case, BLG model can have
gauge symmetry based on an arbitrary Lie algebra ${\bf g}$.
Another option is to lower the supersymmetry to $\cN=6$
where we can have $U(N)\times U(N)$ or $SU(N)\times SU(N)$
gauge symmetry \cite{Aharony:2008ug} 
(see also \cite{Benna:2008zy,Nishioka:2008gz,Honma:2008jd,
Imamura:2008nn,Minahan:2008hf,Gaiotto:2008cg})
which may be alternatively realized by 
Lie 3-algebra which is not anti-symmetric \cite{Bagger:2008se}.

One of the crucial test of multiple M2 brane theory is
whether one can reproduce the celebrated $N^{3/2}$ 
scaling law for entropy \cite{Klebanov:1996un} as is predicted by 
AdS/CFT correspondence.
For any theory based on Lie algebra, however, 
this seems to be difficult. The number of moduli
is related to the rank of the Lie algebra and the number
of the generators is given by the dimension.  It will
produce $N^{3/2}$ scaling only if one consider
delicately chosen  tensor products
of Lie groups \cite{FigueroaO'Farrill:2008zm}
or so far  hidden mechanism changes the 
degrees of freedom.

In this paper, we take a different approach to this issue.
Some time ago, it was proved that BLG model based on
infinite dimensional Lie 3-algebra
defined by Nambu-Poisson bracket is equivalent to M5 brane
world volume theory
\cite{Ho:2008nn, Ho:2008ve} (see also \cite{Li:2008ez, Park:2008qe,
Bandos:2008fr,Furuuchi:2008ki}). 
What we are going to do is to cut-off
this Lie 3-algebra to finite dimensions.  It is actually very natural
to expect to have $N^{3/2}$ law from the following geometrical 
reason.

We note that the Nambu-Poisson bracket is defined by
\cite{Nambu},
\ba
\{f, g, h\} =\sum_{\mu,\nu,\rho=1}^3 \epsilon_{\mu\nu\rho}
\partial_\mu f \partial_\nu g \partial_\rho h.
\ea
Here $f,g,h$ are arbitrary 
functions of three variables $x^1, x^2, x^3$.
Suppose we can truncate 
this infinite dimensional Hilbert
space into a finite dimensional one, let us assume
that we have $N$ degrees of freedom for each dimensions.
The number of independent generators 
behaves as $\#G\sim N^3$.  On the other hand, the number
of M2 branes is, roughly speaking,
 identified with the number of the moduli
which are related to mutually
commuting degree of freedom.  In this case, 
due to the  structure of the Nambu-Poisson
bracket, 
mutually commuting generators
may be taken as functions which 
depend only on two variables, say
$x^1, x^2$.  The number of such generators can be
estimated as $\#M \sim N^2$.  By combining it, we have
the desired scaling $\#G\sim (\#M)^{3/2}$!

In this paper, 
by generalizing the
procedure considered in \cite{Ho:2008bn},
we show how to obtain a finite dimensional Lie 3-algebra from a truncation
of the Hilbert space where Nambu-Poisson bracket is defined.
The fundamental identity of the Lie 3-algebra is preserved
by the cut-off but 
it becomes generally difficult to 
keep a non-trivial invariant metric.
Therefore, although it is difficult to write BLG action,
we can define the $\cN=8$ supersymmetric equation of motion
as considered in  \cite{Gran:2008vi}.
The counting of the moduli is given as above and 
we obtain the $N^{3/2}$ scaling law of entropy rather robustly. 

By definition, our truncated algebra becomes the infinite
dimensional Lie 3-algebra from Nambu-Poisson
bracket in the large $N$ limit.  In this sense,  it gives an intermediate
geometrical structure between M2-brane and M5-brane.  
This is somewhat analogous to the fact that
D $(p+2)$-brane is obtained by collecting large $N$ limit of D $p$-brane.
In this sense, it may serve as a candidate of multiple M2 branes
although it requires many improvements to define a realistic
theory.\footnote{
We note that a derivation of $N^{3/2}$ law 
for M2 branes was considered
previously in \cite{Berman:2006eu} 
(see also \cite{Berman:2007bv,Copland:2007by}) in
the context of Basu-Harvey equation  \cite{Basu:2004ed}
which describes a ``ridge" configuration of
M2-M5 system.  Their analysis is based on
the fuzzy $S^3$ defined in 
\cite{Ramgoolam:2001zx,Ramgoolam:2002wb}.
Since it appeared before \cite{Bagger:2007jr},
the essential ingredients of the BLG model such
as Lie 3-algebra and the fundamental identity were
not taken into account.  
}

For other important results on multiple M2-brane,
see for example \cite{other_papers}.

\section{Truncation of Nambu-Poisson bracket}
We start from a Nambu-Poisson bracket defined
by local coordinates $x^\mu$ ($\mu=1,\cdots, d$) by
\ba\label{NP0}
\{f_1, f_2, f_3\}:= P(f_1, f_2, f_3)
:= \sum_{\mu_1, \mu_2, \mu_3=1}^d
P^{\mu_1 \mu_2 \mu_3}(x) \partial_{\mu_1} f_1
\partial_{\mu_2} f_2 \partial_{\mu_3} f_3
\ea
where $P^{\mu_1 \mu_2 \mu_3}(x) $ is an anti-symmetric tensor.
In order to apply to the BLG model, it is essential
to assume here that the Nambu-Poisson
bracket satisfies fundamental identity,
\ba
&&\{f_1, f_2,\{ f_3, f_4, f_5\}\}=\{\{f_1, f_2,f_3\},f_4,f_5\}\nn\\
&& ~~~~~~~
+\{f_3,\{f_1, f_2,f_4\},f_5\}+ \{f_3, f_4,\{f_1, f_2,f_5\}\}\,.
\label{FI}
\ea
The Leibniz rule,
\ba
\{f_0 f_1, f_2, f_3\}=
f_0\{f_1, f_2, f_3\}+\{f_0, f_2, f_3\}f_1\,,
\label{Leibniz}
\ea
is usually required in the literature.  In the context of BLG model,
the role of this condition is not very clear at this moment.
The fundamental identity imposes a severe constraint on 
$P^{\mu_1 \mu_2 \mu_3}(x)$.  In mathematical literature,
it is known that the fundamental identity implies the decomposability of
$P$ (see for example \cite{Vaisman} and references therein). 
Namely it should be rewritten as
\ba
&&
P=P^{\mu_1 \mu_2 \mu_3}(x)\partial_{\mu_1}\wedge\partial_{\mu_2}\wedge
\partial_{\mu_3}=V_1\wedge V_2\wedge V_3\,,\\
&& V_i(x) = V_i^\mu(x) \partial_\mu\,.
\ea
It implies that the Nambu-Poisson bracket is essentially
defined on three dimensional subspace ($\mathcal{N}$) specified by
the tangent vectors $V_i$ ($i=1,2,3$). 
In \cite{Ho:2008nn,Ho:2008ve}, it was used
to obtain the M5 brane from BLG model whose
world volume is the product $\mathcal{M}\times \mathcal{N}$.
In the following, since we need to restrict 
$P^{\mu_1 \mu_2 \mu_3}(x)$ to be polynomials of fixed degree
for the consistency of the cut-off,
we will not use this decomposability. 
When $P^{\mu_1 \mu_2 \mu_3}(x)$ are homogeneous
polynomial of degree $p$, we call the 3-bracket
as the {\em homogeneous} Nambu-Poisson bracket
\footnote{
Let us briefly mention the previous studies
on the quantum Nambu bracket.
One of the most natural direction is to seek an
analog of the Moyal product as a deformation of
Poisson bracket.  It was studied most
extensively by Takhtajan \cite{Takhtajan} and his
collaborators.  Despite much efforts, however,
the natural analog of the Moyal product has not been
found so far. At some point, they changed the strategy
and found a deformation of Nambu-Poisson bracket
 which was called 
``Zariski quantization" \cite{Dito:1996xr}. 
This construction, however,
needs to use an analog of the 
second quantized operators and is infinite dimensional
by its nature.
Another approach is to use a generalization 
of the matrix commutator (see for example 
\cite{Curtright:2002fd}). Although it gives rise to
a very simple finite dimensional system,
the triple commutator satisfies so called
generalized Jacobi identity instead of the
fundamental identity.  In this sense, it is not
obvious how to apply their algebraic structure
to the BLG model.
The third approach is to use the cubic matrix
(three index object like ``$A_{ijk}$")
to represent the 3-algebra (see for example
\cite{Awata:1999dz, Kawamura:2002yz}).
Although there were some success, for example
in the construction of ``representations"
of $A_4$ algebra \cite{Kawamura:2003cw},
the cubic matrix in general does not satisfy the
fundamental identity.  So it is still mysterious
how to apply it to BLG model.
To summarize, although there are some 
attractive proposals in the quantum 
Nambu bracket, our simple cut-off procedure 
of the Nambu-Poisson seems to be the first example
which can be readily applicable to BLG model.
We do not, of course, mean that other approaches
which we mentioned are meaningless in the BLG model.
On the contrary we are trying to find applications
of  these constructions which we hope to report in the
near future.}.

In \cite{Ho:2008bn}, a truncation of
the Nambu-Poisson bracket (\ref{NP0}) which satisfies
the fundamental identity was proposed.
The idea was to truncate the Hilbert space 
$C(X)$ (functions on $X$) to polynomials of $x^\mu$
of degree $\leq N$.  We will write this truncated
Hilbert space as $C(X)_N$.
For such truncation to work properly,
we need to restrict the anti-symmetric tensor $P^{\mu_1\mu_2 \mu_3}(x)$
to be a homogeneous polynomial of degree $p> 0$.

On $C(X)_N$, we redefine the Nambu-Poisson bracket
to project out all the monomials of order $>N$.
We denote such projector as $\pi_N$ which acts 
on the polynomials of $x^\mu$ as
\ba
&& \pi_N \left(\sum_{n_1,\cdots, n_d=0}^\infty c(n_1,\cdots, n_d)
(x^{1})^{n_1}\cdots (x^{d})^{n_d}\right)\nn\\
&&~~~~~=\sum_{n_1,\cdots, n_d=0}^{|\vec n|\leq N}  c(n_1,\cdots, n_d)
(x^{1})^{n_1}\cdots (x^{d})^{n_d}\,,
\ea
where $|\vec n|:=\sum_{i=1}^d n_i$.
The Nambu-Poisson bracket on the
truncated Hilbert space $C(X)_N$ is then defined as
\ba
\{f_1,f_2,f_3\}_N :=\pi_N\left(P(f_1, f_2, f_3)\right)\,.
\ea
It satisfies
the fundamental identity
\ba
&&\{f_1, f_2,\{ f_3, f_4, f_5\}_N\}_N=\{\{f_1, f_2,f_3\}_N,f_4,f_5\}_N\nn\\
&& ~~~~~~~
+\{f_3,\{f_1, f_2,f_4\}_N,f_5\}_N+ \{f_3, f_4,\{f_1, f_2,f_5\}_N\}_N\,,
\label{FI1}
\ea
because of the following reason \cite{Ho:2008bn}.
For simplicity, we assume $f_i$ to be a monomial
of degree $p_i$.  Since (\ref{FI1}) is satisfied trivially
if $f_i=$const, one may assume $p_i>0$.
The fundamental identity becomes nontrivial if
the outer bracket is non-vanishing, namely,
\ba
p_1+p_2+p_3+p_4+p_5-6+2p\leq N\,.
\ea
The fundamental identity is broken if the inner bracket vanishes
due to the projection.  This does not happen.  For example,
for the left hand side of (\ref{FI1}), the above inequality
together with $p_i\geq 1$ implies
\ba
p_3+p_4+p_5\leq N+6-2p-p_1-p_2
\leq N+4-2p \leq N+3-p\,.
\ea
In the last inequality, we used $p\geq 1$.
Therefore whenever the outer bracket does not vanish, the
value for the outer bracket is identical with the original bracket.
So the FI on the truncated Hilbert space 
comes from the FI on the original space. 

$C(X)_N$ is generated by finite number of monomials,
$(x^{1})^{n_1}\cdots(x^{d})^{n_d}:= T(\vec n)=T(n_1,\cdots, n_d)$ where
$n_i\geq 0$ and $|\vec n| \leq N$. 
The truncated Nambu-Poisson bracket defines 
a Lie 3-algebra,
\ba
\{T(\vec n_1), T(\vec n_2), T(\vec n_3)\}_N
= \sum_{\vec n_4} {f^{\vec n_1 \vec n_2 \vec n_3}}_{\vec n_4} T(\vec n_4)\,,
\label{3-algebra}
\ea
which satisfies the fundamental identity,
\ba
&&{f^{\vec n_3\vec n_4\vec n_5}}_{\vec n_6}
{f^{\vec n_1\vec n_2\vec n_6}}_{\vec n_7}
={f^{\vec n_1 \vec n_2 \vec n_3}}_{\vec n_6}
{f^{\vec n_6\vec n_4 \vec n_5}}_{\vec n_7}
\nn\\
&&~~~~~~~~+
{f^{\vec n_1 \vec n_2 \vec n_4}}_{\vec n_6}
{f^{\vec n_3\vec n_6\vec n_5}}_{\vec n_7}
+
{f^{\vec n_1 \vec n_2 \vec n_5}}_{\vec n_6}
{f^{\vec n_3\vec n_4 \vec n_6}}_{\vec n_7}\,.
\ea
We remark that the geometrical meaning of the algebra 
becomes clear when one takes the large $N$
limit where the algebra of polynomials can be completed in different ways 
and this corresponds to different topological spaces. 

We note that because of the constraint $p\geq 1$,
we cannot define the truncated 3-algebra
from the Jacobian,
\ba
P=\partial_1\wedge \partial_2 \wedge \partial_3\,.
\ea

As for the Leibniz rule (\ref{Leibniz}), 
we have to be careful how to define the product of
functions in the truncated Hilbert space.
We define
\ba
f\bullet_N g=\pi_N(f g)\,,
\ea
which gives a commutative and associative product
on the truncated space\footnote{
This reminds us of the abelian deformation 
of the Nambu-Poisson bracket in \cite{Dito:1996xr}.}.
We replace the Leibniz rule by using this
product rule,
\ba
\{f_0 \bullet_N f_1, f_2, f_3\}_N=
f_0\bullet_N \{f_1, f_2, f_3\}_N+\{f_0, f_2, f_3\}_N\bullet_N f_1\,.
\label{Leibniz2}
\ea
We show that this condition is also satisfied for $p\geq 1$.

Let us assume that $f_i$ are monomials of $x$
with degree $p_i\geq 1$ since the Leibniz rule is
trivially satisfied when $p_0=0$ or $p_1=0$.
The condition that
the left hand side of (\ref{Leibniz2})
is non-vanishing is
\ba
p_0+p_1\leq N\,,\quad
p_0+p_1+p_2+p_3+p-3\leq N\,.
\ea
Since the second condition gives a stronger condition than
the first for $p\geq 1$, we take the second condition.
The first term on the right hand side is non-vanishing if
\ba
p_1+p_2+p_3+p-3\leq N\,,\quad
p_0+p_1+p_2+p_3+p-3\leq N\,.
\ea
Again the second condition gives a stronger constraint.
The second term on the left hand side is non-vanishing
with the same condition.  To summarize, the conditions for
the both sides of equation are the same.  So the
truncation is compatible with the Leibniz rule
(\ref{Leibniz2}) for $p\geq 1$.

\section{Homogeneous Nambu-Poisson brackets
and associated (fuzzy) geometries}

For any homogeneous Nambu-Poisson, we can define
a truncated algebra for each $N$. In the following,
we give some examples of homogeneous algebra
which satisfies the fundamental identity and
associate each algebra with a 
three dimensional manifold. In general,
we have descriptions of the homogeneous
Nambu-Poisson in terms of $d$ variables.
The fact that Nambu-Poisson bracket is defined
in 3-dimensions can be derived by observing that
there are $d-3$ elements $f_a(x)$ which commute
with any functions of $x$, namely,
\ba
\{f_a, g, h\}=0,\quad \mbox{for any }g, h\,.
\ea 
So one may use the hyper-surface defined by
$f_a(x)=c_a$ ($a=1,\cdots, d-3$) 
as the definition of 3 dim submanifold
in $\mathbf{R}^d$.
If we introduce the cut-off, one may call the corresponding
geometry as ``fuzzy spaces" by employing the terminology
of the noncommutative geometry although
our definition of the deformation is
very different.

We start from the $p=1$ case.
In this case, we call the bracket as  {\em linear}
Nambu-Poisson bracket \cite{DZ} in the following. 
We note that the coordinates $x^\mu$
define a Lie 3-subalgebra,
\ba
\{x^{\mu_1}, x^{\mu_2}, x^{\mu_3}\}
=\sum_{\mu_4}
{f^{\mu_1\mu_2 \mu_3}}_{\mu_4} x^{\mu_4}\,,
\quad
P^{\mu_1 \mu_2 \mu_3}(x)=\sum_{\mu_4} 
{f^{\mu_1\mu_2 \mu_3}}_{\mu_4} x^{\mu_4}\,.
\ea
The mathematical classification of the linear
Nambu-Poisson was already made
and it was reviewed in \cite{Ho:2008bn}.
It is classified into two groups,

\paragraph{Type I: } For each $-1\leq r\leq 3$, 
$0\leq s\leq  \min(3-r,d-4)$ 
one may define the bracket as
\ba
&& P^I_{(r,s)}=\sum_{j=1}^{r+1} \pm x^j \partial_1\wedge
\cdots \backslash\!\!\!\!\partial_j \cdots \wedge
\partial_4\nn\\
&&~~~~~~~~~~~~~+\sum_{j=1}^s \pm x^{n+j+1}\partial_1\wedge
\cdots \backslash\!\!\!\!\partial_{r+j+1}
\cdots\wedge\partial_{4}\,.
\ea
Here $\backslash\!\!\!\!\partial$ means that we delete that
element in the wedge product.
\paragraph{Type II: }
\ba
P^{II}_a=\partial_1\wedge\partial_2\wedge(\sum_{i,j=3}^d
a_{ij} x^i\partial_j)\,.
\ea
For type I case, we can choose the plus/minus sign for each term
in the summation.

In the following, we pick up interesting examples
that come from this classification theorem for each $d$,
the number of coordinates.
 
\paragraph{$d=3$ : }

The only possibility comes from the type II algebra,
\ba\label{s1r2}
P= \partial_1\wedge\partial_2\wedge x^3\partial_3\,.
\ea
In this case, 
the $x^3$ may be taken as a real number or a phase $e^{i \theta_3}$.
When $x^3$ is taken as real, and with an appropiate completion, 
the truncated algebra can be thought as a deformation of $\mathbf{R}^3$ 
 \footnote{
To avoid possible confusion, we emphasis that this is not the standard $R^3$
as a Poission manifold. There  the
Poisson structure is $SO(3)$ and translationally invariant.
}. 
Due to the extra factor of $x^3$, the Poisson structure \eq{s1r2} 
breaks $O(3)$ symmetry. In the
correspondence with M5 brane
\cite{Ho:2008nn,Ho:2008ve}, $P$ represents
the 3-form flux on M5 world volume.  The 
breakdown of rotational symmetry comes from the fact that the
3-form background does not respect the symmetry.
When $x^3$ is a phase, one can think of the truncated algebra as a
deformation of $ \mathbf{R}^2\times \mathbf{S}_{+}^1$, where
$\mathbf{S}_{+}^1$ is dual to the algebra of functions with only
non-negative Fourier modes. In this case 
$P \sim \partial_1\wedge\partial_2\wedge\partial_{\theta_3}$ defines a
Nambu-Poisson bracket on $\mathbf{R}^2\times \mathbf{S}_+^1$.


\paragraph{$d=4$ :}
In this case a variety of examples come from
type I.  For $r=3,s=0$ case, a well known example is
\ba
\label{s3}
P=x^1\partial_2\wedge\partial_3\wedge \partial_4
-x^2\partial_1\wedge\partial_3\wedge \partial_4
+x^3\partial_1\wedge\partial_2\wedge \partial_4
-x^4\partial_1\wedge\partial_2\wedge \partial_3\,.
\ea
In this case, the 3-algebra generated by the coordinates
is $A_4$. It defines a Nambu-Poisson bracket on $\mathbf{S}^3$
since $r^2=(x^1)^2+(x^2)^2+(x^3)^2+(x^4)^2$ becomes the
center of the 3-algebra.  Namely,
\ba
P(r^2 f_1, f_2, f_3)= r^2 P(f_1, f_2, f_3)\,,
\ea
for any $f_1, f_2, f_3$. So one may put $r^2=\mbox{const}$.
The truncated algebra defines a 
fuzzy $\mathbf{S}^3$ in $\mathbf{R}^4$.

{}From this example, by taking Wick rotation, we obtain other
examples.  For example, the bracket after $x^4\rightarrow i x^4$,
\ba
P=x^1\partial_2\wedge\partial_3\wedge \partial_4
-x^2\partial_1\wedge\partial_3\wedge \partial_4
+x^3\partial_1\wedge\partial_2\wedge \partial_4
+x^4\partial_1\wedge\partial_2\wedge \partial_3\,,
\ea
defines a bracket on $\mathbf{dS}^3$ since 
$(x^1)^2+(x^2)^2+(x^3)^2-(x^4)^2$ becomes the
center of the algebra and can be set to a constant.

Similarly after taking the Wick rotation for $x^3, x^4$,
we obtain
\ba
P=x^1\partial_2\wedge\partial_3\wedge \partial_4
-x^2\partial_1\wedge\partial_3\wedge \partial_4
-x^3\partial_1\wedge\partial_2\wedge \partial_4
+x^4\partial_1\wedge\partial_2\wedge \partial_3\,.
\ea
In this case, $(x^1)^2+(x^2)^2-(x^3)^2-(x^4)^2$
becomes the center of 3-algebra and can be set
to a constant which defines
$\mathbf{AdS}^3$.

For $r=2$, $s=0$, we have
\ba
P_{(2,0)}&=& x^1\partial_2\wedge\partial_3\wedge \partial_4
+ x^2\partial_1\wedge\partial_3\wedge \partial_4
\pm x^3\partial_1\wedge\partial_2\wedge \partial_4\nn\\
& =&( x^1\partial_2\wedge\partial_3
+ x^2\partial_1\wedge\partial_3
\pm x^3\partial_1\wedge\partial_2)\wedge \partial_4\,.
\ea
The center takes the form $(x^1)^2+(x^2)^2\pm (x^3)^2$
and 3d manifold associated with it is $\mathbf{S}^2\times \mathbf{R}$
or $\mathbf{(A)dS}^2\times \mathbf{R}$
where $\mathbf{R}$ is described by $x^4$. 
For finite $N$, we have a deformation of these manifold. 

In order to have $s>0$, we need to take $d>4$.
For example for $s=1$, we need $d=5$ and
\ba
P_{2,1}=P_{2,0}\pm x^5 \partial_1 \wedge \partial_2 
\wedge \partial_3\,.
\ea
In this case, since $x^5$ does not appear in the
derivative, it is the center of 3-algebra.
Actually the algebra for the linear functions is
identical with the Lorentzian algebra
\cite{Gomis:2008uv,Benvenuti:2008bt,Ho:2008ei} 
for $g=SU(2)$ or $SL(2)$
where $x^4, x^5$ play the role of $T^0, T^{-1}$
respectively.  In general the parameter $s$ represents
the number of pairs of the Lorentzian generators.
For smaller $r$ we can add more pair ($3-r$) of Lorentzian generators.
For $r=2, s=1$, the center of the algebra becomes
\ba
(x^1)^2+(x^2)^2\pm (x^3)^2\pm 2 x^4 x^5,\ 
\mbox{and}\ x^5\,,
\ea
to which we can assign arbitrary value.

For $r=1$ we obtain $\mathbf{S}^1\times \mathbf{R}^2$
or $\mathbf{R}^3$
and its generalizations with pairs of Lorentzian generators.
We note that here we obtain $\mathbf{S}^1$ 
or $\mathbf{R}^1$ from a
constraint $(x^1)^2\pm (x^2)^2=$const.
For $r=0$, we obtain $\mathbf{R}^3$ with the bracket,
\ba
P=x^1 \partial_2 \wedge \partial_3 \wedge \partial_4\,.
\ea
Here $x^1$ becomes the center of 3-bracket and
can be set to a constant. 

For $r=-1$, we have only the Lorentzian pairs.

For $p>1$, we do not have the classification theorem.
We have, however, a few interesting examples
of Nambu-Poisson bracket where fundamental identity is 
satisfied.

For $p=2$, we have, for example,
\ba\label{s1s1r1}
P= \partial_1\wedge x^2\partial_2 \wedge
x^3\partial_3.
\ea
If we take $x^{2,3}$ real, the we have a deformed $\mathbf{R}^3$ with 
linear flux introduced in two directions.  By taking $x^2$ or/and $x^3$
to be a phase, we can also have deformed $\mathbf{R}^2 \times
\mathbf{S}_+^1$ or $\mathbf{R} \times
\mathbf{T}_+^2$ 
($\mathbf{T}_+^2$ represents $\mathbf{S}_+^1 \times \mathbf{S}_+^1$).

Another example is
\ba
P=(\epsilon_{\mu\nu\lambda}x^\mu\partial_\nu
\wedge \partial_\lambda)\wedge x^4\partial_4
\ea
which 
can describe deformation of 
$\mathbf{S}^2\times \mathbf{R}^1$ or $\mathbf{S}^2\times
\mathbf{S}^1_+$.

For $p=3$, we have an example,
\ba\label{t3}
P=x^1\partial_1\wedge x^2\partial_2 \wedge
x^3\partial_3
\ea
which can describe deformed $\mathbf{R}^3$, $\mathbf{R}^2\times
\mathbf{S}^1_+$, $\mathbf{R}\times \mathbf{T}^2_+$ or $\mathbf{T}^3_+$
depending on the interpretation of $x^\mu$.

This last example will be used in the following 
since it has the simplest structure.  In particular,
the algebra (\ref{3-algebra}) takes the following form
(after minor change of the normalization factors),
\ba
\{ T(\vec n_1), T(\vec n_2), T(\vec n_3) \} 
= \vec n_1\cdot (\vec n_2 \times \vec n_3) 
T(\vec n_1+\vec n_2+\vec n_3)\,.
\ea
The truncated version becomes
\ba\label{t3tr}
\{T(\vec n_1), T(\vec n_2), T(\vec n_3)\}_N
= \vec n_1\cdot (\vec n_2 \times \vec n_3) 
\theta\left(N-|\sum_i \vec n_i|\right)
T(\vec n_1+\vec n_2+\vec n_3)
\ea
where $(\vec n_i)_j\geq 0$ 
and
\ba
\theta(n)=\left\{
\begin{array}{ll}
1\quad & n\geq 0\\
0 \quad & n<0
\end{array}
\right.\,.
\ea
The explicit form of the algebra for other cases is 
straightforward to write down.  For example,
$\mathbf{S}_3$ case eq.(\ref{s3}) is given as
\ba
\{ T(\vec n_1), T(\vec n_2), T(\vec n_3)\}=
\epsilon_{\mu\nu\lambda\rho}(n_1)_\nu (n_2)_\lambda (n_3)_\rho
T(\vec n_1+\vec n_2+\vec n_3-\vec \sigma+2\vec e_\mu),
\ea
where
$(\vec e^\mu)_\nu=\delta_{\mu\nu}$ and
$\vec\sigma=\sum_{i=1}^4 \vec e_\mu
$.  The truncated 3-algebra can be obtained by restricting
the generators to $|\vec n|\leq N$ and including
a truncation factor $\theta(N+2-\sum_i |\vec n_i|)$ on
the right hand side.

\section{Application to BLG model and counting entropy}
As we show in the appendix, the metric of the truncated
Nambu-Poisson bracket has a trivial structure
and is useless in the construction of the invariant
Lagrangian\footnote{
Of course, there may be a chance to add extra generators
to obtain a nontrivial and useful  metric 
as in \cite{Gomis:2008uv, Benvenuti:2008bt, Ho:2008ei}.}.
Nevertheless, we can write down an N=8 supersymmetric
equation of motion in terms of the structure constants
of the Lie 3-algebra which satisfies the fundamental identity
\cite{Gran:2008vi},
\ba
&& D^2 X^I_A-\frac{i}{2}\bar\Psi_C \Gamma^I_J X^J_D \Psi_B{f^{CDB}}_A
+\frac{1}{2} {f^{BCD}}_A {f^{EFG}}_D X^J_B X^K_C X^I_E X^J_F X^K_G =0,
\;\;\;\;\;\;\  \\
&&\Gamma^\mu D_\mu \Psi_A +\frac{1}{2}\Gamma_{IJ} 
X^I_C X^J_D \Psi_B {f^{CDB}}_A=0, \\
&& {(\tilde F_{\mu\nu})^B}_A+\epsilon_{\mu\nu\lambda}
(X^J_C D^\lambda X^J_D+\frac{i}{2}\bar\Psi_C \Gamma^\lambda\Psi_D)
{f^{CDB}}_A=0\,.
\ea
The SUSY transformation is
\ba
&& \delta X^I_A =i\bar\epsilon \Gamma^I \Psi_A\\
&&\delta\Psi_A=D_\mu X^I_A \Gamma^\mu\Gamma_I\epsilon
-\frac{1}{6}X^I_B X^J_C X^K_D {f^{BCD}}_A \Gamma_{IJK}\epsilon\\
&& \delta(\tilde A_\mu)^B_A=i\bar\epsilon \Gamma_\mu\Gamma_I X^I_C\Psi_D
{f^{CDB}}_A\,.
\ea
An essential point here is that the structure constant
contracted with metric $f^{ABCD}={f^{ABC}}_E h^{ED}$
does not appear at all.
It enables us to discuss important issues such as
the BPS equation or the moduli
without knowing the Lagrangian.

Let us pick the algebra (\ref{t3}) and study the
moduli.  From the equation of motion,
the moduli would be described by solutions of the equation
\ba\label{moduli}
{f^{EFG}}_D X^I_E X^J_F X^K_G=0\,.
\ea

We have to be careful in the structure of
the truncated algebra.  In the appendix, 
we show that the algebra (\ref{t3})
has a structure which is similar to the
Lorentzian algebra \cite{Gomis:2008uv,Benvenuti:2008bt,Ho:2008ei}.
Namely after removing generators which decouple from
the algebra, the set of generators is classified into
three subsets.  If we use a notation
similar to \cite{Ho:2008ei},
 (i) $\mathcal{A}'_0$: the generators which do not
 appear on the right hand side of 3-commutator,
 namely the generator $T^D$ where ${f^{ABC}}_D=0$
 for any $A,B,C$. Such generators 
have the form $T(\vec k)$ where one or two components
 of $\vec k$ are zero. (ii) $\mathcal{A}'_{-1}$: the generators
 which are in the center of 3-algebra. Namely the generator
 $T^A$ where ${f^{ABC}}_D=0$ for any $B,C,D$.
Such generators
take the form $T(\vec k)$ where $\sum_i k_i=N-1, N$.
(iii) $\hat\mathcal{A}$: generators which do not belong
to $\mathcal{A}'_0$ nor $\mathcal{A}'_{-1}$.
The difference from \cite{Gomis:2008uv,Benvenuti:2008bt,Ho:2008ei}
is that we have a large number ($O(N^2)$) of elements
in $\mathcal{A}'_0$ and $\mathcal{A}'_{-1}$.

The roles of fields in each subgroup are similar
to \cite{Gomis:2008uv,Benvenuti:2008bt,Ho:2008ei}.
Let us denote the generic fields which belong to 
$\mathcal{A}'_0$, $\mathcal{A}'_{-1}$, $\hat\mathcal{A}$
as $X$,  $Y$, $Z$ respectively.
Then the equation of motion is written schematically as
\ba\label{eom1}
\partial^2 X=0, \quad
\partial^2 Y=F_1(X,Z),\quad
\partial^2 Z=F_2(X,Z)\,,
\ea
and SUSY (and gauge) transformations are written similarly,
\ba\label{trans1}
\delta X=0,\quad
\delta Y= G_1(X,Z),\quad
\delta Z=G_2(X,Z)\,,
\ea
where $F_{1,2}, G_{1,2}$ represent some nonlinear functions.
To find moduli, we can put the left hand side of equation of motion
(\ref{eom1}) to be zero. 

First we note that there is no constraint for $Y$ from (\ref{moduli}).  
Besides, $Y$ fields
never appear in the nonlinear terms in the equations of motion. 
We can take any solutions of $Y$ of their equations of motion, 
and it will not have any effect on the rest of the fields. 
In this sense, the $Y$ fields should be viewed 
as non-physical fields, and we will not treat them as part of 
the moduli. 
\footnote{
On the other hand, if we treat them as part of the moduli, 
the number of solutions of (\ref{moduli}) can be of order $N^3$. 
We can take 6 of the scalars $X^I$ to be $Y$ fields, 
and the rest 2 of the $X^I$'s can be arbitrary. 
For large $N$, the number of free parameters in the 2 arbitrary 
fields $X^I$ dominates and it is proportional to $N^3$.}

Secondly, if we assign VEV to $X$, the field equation and
the symmetry transformations do depend
on the VEV.  On the other hand,
the SUSY (gauge) transformation (\ref{trans1})
for $X$ implies that these symmetries are not
violated.  
This behavior is what one expects for a vacuum state. 
On the other hand, in the Lorentzian BLG model \cite{Ho:2008ei}, 
the VEV for $X^I_0$ was interpreted as
the coupling constant of the super Yang-Mills
theory on D2 and hence is not counted as part of the moduli space. 
Further analysis is needed 
to decide whether these are to be counted as part of
the moduli space or not.  However we will see that including them or 
not does not affect our entropy counting below. 

Finally the assignment of VEV for $Z$
does not seem to have such strange behavior.
Therefore, this is the degree of freedom which
should be  identified with the moduli of M2 brane
in ordinary sense.

It turns out that the equation (\ref{moduli})
can give rise to various solutions.
For the 3-algebra (\ref{t3}),
three polynomials $f_1,f_2,f_3$ which depend only on
two polynomials of $x$, say $g_1(x), g_2(x)$
in general commute with each other,
\ba
\{ f_1(g_1, g_2), f_2(g_1, g_2), f_3(g_1, g_2)\}_N=0\,.
\ea 
Therefore the moduli space is described by
(truncated) polynomials of $g_1(x)$ and $g_2(x)$.
Depending on the choice of $g_{1,2}$,
we have different type of ``Higgs" branches.

If we take both $g_{1,2}$ as function of
single variables, say $g_1=x^1$, $g_2=(x^2)^m$,
all the functions of $g_{1,2}$ belong to 
the group $\mathcal{A}'_0$.
The number of such functions is of the order of $N^2$. 
As we explained above,
these may or may not be counted as part of the moduli space.

On the other hand, suppose we take 
$g_{1,2}$ such that their polynomials depend 
on all the coordinates non-trivially,
for example $g_{1}=x^1+x^2$ and $g_2=(x^3)^2$,
the set  of polynomials of them contains
 elements belonging to $\hat\mathcal{A}$.
In this case, the VEVs are assigned to $Z$ fields and
should be interpreted as the moduli of M2 branes. 
We can count the number of the M2 branes for
given set of $g_{1,2}$.  Suppose we choose them such that
all the VEVs of fields can be interpreted as the moduli
of M2 branes.  If the degree of $g_{1,2}$ is
$n_{1,2}$ respectively, 
the number of independent generators are approximately
$\frac{N^2}{2 n_1 n_2} \sim N^2$ as long as $n_{1,2}$ are much smaller than $N$.
We have the estimate for the number of membrane as
\ba
\#M\sim N^2.
\ea
This permits us to calculate the behavior of the entropy.
The number of fields 
is given as the number of generators ($\# G$).
It can be estimated as
\ba
\# G= \frac{(N+1)(N+2)(N+3)}{6} \sim N^3/6\sim (\# M)^{3/2} . 
\ea
This is the celebrated $N^{3/2}$ law for M2-brane.

One may do essentially the same counting for other
$d=3$ algebras associated with $\mathbf{R}^3$
(\ref{s1r2},\ref{s1s1r1}) which give the same
behavior.
So one may guess the behavior of  $N^{3/2}$ law
as a generic feature of the $d=3$ truncated
Nambu-Poisson 3-algebras.

We note that there are some subtlety if one continues to do the 
similar analysis for $d>3$ cases.  As we have seen,
there are $d-3$ generators $\phi_s(x)$ which 
satisfy,
\ba
\{\phi_s f_1, f_2, f_3\}=\phi_s\{f_1, f_2, f_3\}
\ea
for any $f_1, f_2, f_3$. One may set such generators
as constant $\phi_s(x)=c_s$ and this constraints gives
3 dimensional algebra.

For the truncated algebras, since such $\phi_s$ has nontrivial
degree as the polynomial of $x$. For example
$\phi=(x^1)^2+(x^2)^2+(x^3)^2+(x^4)^2$ which appear
for $\mathbf{S}^3$ case has degree two.
So the above relation should be modified as
\ba
\{\phi_s \bullet_N f_1, f_2, f_3\}_N =\phi_s\bullet_N\{f_1, f_2, f_3\}_{N-|\phi_s|}
\ea
where $|\phi_s|$ is the degree of  $\phi_s$.
It implies that we cannot put $\phi_s$ to a c-number
if we want to keep the fundamental identity.
If we treat them as the independent generators, we would
have different scaling.  For example, for any $d=4$ cases,
we have a simple estimate that
\ba
\# M\sim O(N^3),\quad
\# G\sim O(N^4)\,.
\ea
Therefore we obtain $N^{4/3}$ relation between
the number of membranes and the number of degree of freedom.
This strange behavior for $d>3$ signals the 
breakdown of the truncation process which does not
properly respect the local factorization of the space
into $3$ dimensional and $d-3$ dimensional spaces. Therefore, this
anomalous scaling law should be understood as
coming from an incorrect regularization of the system.

\section{Discussion}

In this paper, we proposed a series of Lie 3-algebra
which has two remarkable properties,
\begin{itemize}
\item $N^{3/2}$ scaling of M2 branes 
with clear geometrical meaning.
\item M5 brane theory in the large $N$ limit 
\cite{Ho:2008nn, Ho:2008ve}.
\end{itemize}
On the other hand, it has obvious shortcomings
at this moment, namely we cannot define non-trivial
Lagrangian with the current form of the algebra.
A hope is that one may cure it by adding some
extra generators as in \cite{Gomis:2008uv,Benvenuti:2008bt,Ho:2008ei}.

Of course, the cut-off algebra which we considered here
is rather exotic algebra which was not considered
seriously in the literature.  For example it would be much more
desirable to do similar truncation by some generalization
of the Moyal product or by some generalization of the concept of
matrices. We note that, however, 
our derivation of $N^{3/2}$ law
is quite robust and the derivation of the scaling law
will be similar even for these cases.

\section*{Acknowledgment}

P.-M. H., Y. M. and S. S. appreciate partial financial support from
Japan-Taiwan Joint Research Program
provided by Interchange Association (Japan)
by which this collaboration is made possible.

The authors thank Kazuyuki Furuuchi, Yosuke Imamura, 
Darren Sheng-Yu Shih, Douglas Smith, 
Wen-Yu Wen and Tamiaki
Yoneya for helpful discussions. 
P.-M. H. is grateful to Anna Lee for assistance in many ways. 
C.-S.C. acknowledges EPSRC for an advanced research fellowship and partial support 
by STFC.  
The work of P.-M. H. is supported in part by
the National Science Council,
the National Center for Theoretical Sciences, 
and the LeCosPA Center at National Taiwan University. 
Y. M. is partially supported by KAKENHI (20540253).

\appendix

\section{Some details on the truncated Nambu-Poisson
algebra (\ref{t3})}

\subsection{Structure of algebra}
We note that the truncated 3-algebra 
on  (\ref{t3})
can be decomposed into three subspaces:
\begin{itemize}
\item[$\mathcal{A}_0$] A subspace spanned by generators 
$T(\vec k)$ where one or two components of $\vec k$ is zero.
In the definition of NP bracket, we always multiply
$x_1 x_2 x_3$ after taking the derivation.  So the generators
which belong to $\mathcal{A}_0$ never appear on the right
hand side of the commutator. We will denote generic generator
which belongs to $\mathcal{A}_0$ as $T_X$.

\item[$\mathcal{A}_{-1}$] A subspace spanned by generators
$T(\vec k)$ where $|\vec k|=N-1, N$.  These generators 
are the center of the algebra, namely 
\ba
\{T_Y, T(\vec p), T(\vec q)\}_N =0 ,
\quad \mbox{for }\forall\,  \vec p, \vec q. 
\ea
where $T_Y$ is a generic generator
which belong to $\mathcal{A}_{-1}$.
It comes the fact that 
we need $|\vec p|,|\vec q|\geq 1$ to have nonvanishing 3-commutator.
These generators can show up
on the right hand side of the 3-bracket.
\item[$\hat\mathcal{A}$] The generators which belong
to neither  $\mathcal{A}_0$ nor $\mathcal{A}_{-1}$.
We will write generic elements of $\hat\mathcal{A}$ as $T_Z$.
\end{itemize}

We note that there are some elements which belong
to $\mathcal{N}=\mathcal{A}_0 \cap \mathcal{A}_{-1}$.
Since every element in this subspace has vanishing
commutator with anybody else and never appears on
the right hand side of the commutator,
they decouple from the algebra as $T(\vec 0)$. 
Therefore, we have to remove them from the algebra.
We will write,
\ba
\mathcal{A}'_0=\mathcal{A}_0/\mathcal{N},\quad
\mathcal{A}'_{-1}=\mathcal{A}_{-1}/\mathcal{N},
\ea
to represent the relevant part of the algebra.
The number of generators which belong to each subspace is
\ba
\#(\hat\mathcal{A})\sim\frac{N^3}{6},\quad
\#(\mathcal{A}_0)\sim\frac{3N^2}{2},\quad
\#(\mathcal{A}_{-1})\sim N^2,\quad
\#(\mathcal{N})\sim 6N\,.
\ea
In the large $N$ limit, the number of the elements which belong
to $\mathcal{A}_0,\mathcal{A}_{-1}$ is large ($O(N^2)$)
but it is still much smaller than that of
$\hat\mathcal{A}$.


\subsection{Invariant metric}
For any element $T_Y^{a}\in  \mathcal{A}'_{-1}$ and any elements
$T_Z^b \in \hat\mathcal{A}$, they must appear on the
right hand side of 3-commutator.  It implies
\ba
\langle T_Y^a, T_Y^b\rangle =\langle [T^P, T^Q, T^R], T_Y^b\rangle
=-\langle T^R, [T^P, T^Q, T_Y^b]\rangle =0\nn\\ 
\langle T_Z^a, T_Y^b\rangle =\langle [T^P, T^Q, T^R], T_Y^b\rangle
=-\langle T^R, [T^P, T^Q, T_Y^b]\rangle =0\nn
\ea
for some $T^P, T^Q, T^R$.
So elements in $\mathcal{A}'_{-1}$ must be orthogonal to 
any elements in $\mathcal{A}'_{-1}$ and $\hat\mathcal{A}$.

Similarly, for two elements in $\mathcal{A}'_{0}$, since they
do not show up in the commutator, there are no constraint
for their inner product from the symmetry:
\ba
\langle T_X^a, T_X^b\rangle = K_{ab} \quad \mbox{(arbitrary)}\,.
\ea

We can also deduce that any elements in $\mathcal{A}'_{-1}$
and $\hat\mathcal{A}$ are orthogonal with the elements of
$\mathcal{A}'_{0}$,
\ba
\langle T_X^a, T_Y^b\rangle= \langle T_X^a, T_Z^b\rangle =0. 
\ea
A proof is as follows.  For the generic elements 
$T_{k_1k_2k_3}\in \mathcal{A}'_{-1}\cup \hat\mathcal{A}$,
we have $k_1,k_2, k_3\neq 0$.
So one may write it as a triple commutator,
\ba
T_{k_1k_2k_3}=\frac{1}{k_1k_2k_3}[T_{k_100}, T_{0k_20}, T_{00k_3}]\,,
\ea
where $T_{k_1k_2k_3}:=T(k_1 \vec e_1+k_2 \vec e_2+k_3 \vec e_3)$.
On the other hand, any element 
$T_{p_1 p_2 p_3}\in \mathcal{A}'_{0}$, one of $p_i$ must be zero.
Let us take it $p_1=0$.  Then we have
\ba
\langle T_{0p_2p_3}, T_{k_1k_2k_3}\rangle& \propto&
\langle T_{0p_2p_3}, [T_{k_100}, T_{0k_20}, T_{00k_3}]\rangle
\nn\\
&=& -\langle [T_{0p_2p_3},T_{0k_20}, T_{00k_3}], T_{k_100}\rangle=0\,.
\ea

Finally for any two elements $T_{p_1p_2 p_3}, T_{q_1 q_2 q_3}$
in $\hat\mathcal{A}$,
one can derive similarly,
\ba
\langle T_{p_1 p_2 p_3}, T_{q_1 q_2 q_3}\rangle &\propto &
\langle T_{p_1p_2p_3}, [T_{q_100}, T_{0q_20}, T_{00q_3}]\rangle\nn\\
&=& -\langle [T_{p_1p_2p_3}, T_{0q_20}, T_{00q_3}],
T_{q_100}\rangle\,.
\ea
On the right hand side, $ [T_{p_1p_2p_3}, T_{0q_20}, T_{00q_3}]$
is zero or belong to either $\hat\mathcal{A}$ or $\mathcal{A}'_{-1}$.
Since the inner product between $\hat\mathcal{A}$ or $\mathcal{A}'_{-1}$
with any element in  $\mathcal{A}'_{0}$ is already shown to be zero, we arrive
at
\ba
\langle T_Z^a, T_Z^b\rangle =0\,,\quad
\forall \ T_Z^{a}, T_Z^{b}\in  \hat\mathcal{A}\,.
\ea 

As we can see, the requirement of invariance imposes very severe
constraints on the form of the metric and at the end the metric has lots
of null directions, making it not useful for physical applications.
The potential term of the BLG model, 
$\langle [X^I, X^J, X^K], [X^I, X^J, X^K]\rangle$ for example, 
is identically zero, 
because 
nontrivial metric components only exist for 
elements in $\mathcal{A}'_0$, 
while elements in $\mathcal{A}'_0$ never appear 
as the result of a 3-bracket.


\end{document}